Scientific Research Publishing

# On a Generic Security Game Model

# Vivek Shandilya[1], Sajjan Shiva[2]

[1]Department of Computing Sciences, Jacksonville University, Jacksonville, USA
[2]Department of Computer Science, University of Memphis, Memphis, USA
Email: shandilya@ju.edu, sshiva@memphis.edu

## Abstract

To protect the systems exposed to the Internet against attacks, a security system with the capability to engage with the attacker is needed. There have been attempts to model the engagement/interactions between users, both benign and malicious, and network administrators as games. Building on such works, we present a game model which is generic enough to capture various modes of such interactions. The model facilitates stochastic games with imperfect information. The information is imperfect due to erroneous sensors leading to incorrect perception of the current state by the players. To model this error in perception distributed over other multiple states, we use Euclidean distances between the outputs of the sensors. We build a 5-state game to represent the interaction of the administrator with the user. The states correspond to 1) the user being out of the system in the Internet, and after logging in to the system; 2) having low privileges; 3) having high privileges; 4) when he successfully attacks and 5) gets trapped in a honeypot by the administrator. Each state has its own action set. We present the game with a distinct perceived action set corresponding to each distinct information set of these states. The model facilitates stochastic games with imperfect information. The imperfect information is due to erroneous sensors leading to incorrect perception of the current state by the players. To model this error in perception distributed over the states, we use Euclidean distances between outputs of the sensors. A numerical simulation of an example game is presented to show the evaluation of rewards to the players and the preferred strategies. We also present the conditions for formulating the strategies when dealing with more than one attacker and making collaborations.

## Keywords

Security System Architecture, Stochastic Game, General Sum Game, Imperfect Information, Numerical Simulation

## 1. Introduction

Ensuring security of a system delivering important services, with exposure to the Internet, has many challenges. When such a system is distributed with many stakeholders , the task of securing such system becomes more complex. A typical system, delivering one or more services would be operating with the co-ordination of processes and data on many different systems exposed to the Internet with different degrees of security. Each stakeholder has his own computational space. Thus each stakeholder providing a service to other stakeholders has to ensure the security of the legitimate users' processes and data while also dealing with malicious users appropriately [1]. In the recent time, many successful attacks have been new, persistent and evolved over time.There have been attempts to employ game theory to model and address such attacks [2]. These attacks are accomplished with multiple steps. When an attacker begins to attack, at each step he is breaching the security of the computational space of his service provider and other related stakeholders. The administrators and security systems must defend the system from such attackers. To do so effectively, it is important to detect each anomalous behavior at the earliest and respond appropriately to minimize the damage. But the information received from the detection of an anomalous behavior may not be sufficient to classify the user as an attacker or a legitimate user who made an innocent fumble or an unusual workload. In such cases the security system must engage with the user to get sufficient information at the earliest to deal with him appropriately. Since such engagement by the administrator or the security system is an expensive overhead, only a detection of an anomalous behavior must trigger such engagement. Thus, we need an event (*detection of anomalous behavior*) driven engagement and a security system facilitating it.

The engagement of the suspected user and the administrator/security system is essentially an interaction between two decision making entities. We observe that game theory provides the relevant framework for such a mechanism design. Hence, we present here an architecture for such a security system based on game theory.

Applying game theory to computer/cyber/network security has been an active research area since 2005 [3]. The crucial part of these attempts is to build appropriate game models to address the security situations. Most of the efforts have been to address a very specific attack by devising a suitable game model and executing a game. Roy, *et al.* [4] provides a survey of the works to model the interaction between a user and a network administrator and a classification of these works based on the game models used. Since our architecture is aimed at addressing different attacks it must have game models appropriate to each attack. But we observe that a generic game model which is abstracted enough to address the typical scenarios would facilitate well in deriving the specific game models. We present a generic game model, a game and numerical simulation to validate our analyses of its utility.

As one of the early attempts, [3] modeled the interaction between an attacker and the administrator as a stochastic game with 14 states considering 3 types of attacks. Their game assumed perfect information. [5] presented a two state, imperfect information, zero sum, stochastic game with numerical simulation showing the advantage of considering the imperfect information. The main motivation for considering the imperfection in the information was the errors in the player's sensors. The error in the sensor makes the player believe that he may be in the states other than the state he really is. We make two extensions to this work. The simulation method used to study and explore the value of the game and the rewards for the strategies are an extension of [6] which has only 5 strategic options.

Our model has the same structure as [5] with one extension. When there are more than two states in the game, this error in sensor could make him mistakenly believe he is in any of the other states than the one he really is in. The error in perceiving the *current state* gets distributed over the other states depending on the sensor reading's error needed to misread the state as the current state. Based on this extension in the game model, a game with five states is designed. The error in the perception of current state gets distributed over the four states other than the real state. Thus, the probability of the player[1] being deluded to be in any of the other states is proportional to the distance between the sensor readings of state defining variables of the other states from that of the current real state.

The second extension is in the game. Since the set of actions available at each state are distinct in the game considered here the extended action sets will be different from the original action sets. For the administrator the sensor error is the cause of perceiving the current state as an information set instead of a single state. The user, who is assumed in this work to have an error free sensor, also faces deception while game enters the Trap state. Then administrator's ploy in the honey pot makes the user unsure of his current state and causes extension of his perceived action set. This increase in the size of the perceived action set at honey pot for the user, making his/her choosing the action less obvious due to increased apparent choices. Following these interactions, administrator will be able to determine the motive of the user under suspicion. This whole interaction based on the administrator ploy is abstracted into one single state.

We analyze the game for its solution discussing the preferred strategies. A numerical simulation implemented in the programming language C of the game validating the analysis is presented. In evaluating the simulation, we discern that the strategy has to precipitate into game theoretic actions, which in turn into computer/network administrator actions. We present a gradation of aggression the players can assert and study the game's dynamics.

The main motivation for this work is to devise a game model and a game generic enough to incorporate real security games, needed to build a security system. [7] presents the architecture to construct a security system based on the game theory approach. The generic game model and the game here are useful to derive different game models and instantiate games, appropriate for different

---

[1]Administrator, as the sensor of the user is considered to be error free we are considering here the worst case scenario for the administrator. Our work can be extended to the case with user too having erroneous sensors.

security situations. The major contributions of this work are summarized as below.

1) We present a security system architecture, its components and the detailed operational flow.

2) We extend the previously proposed game model by devising a method for modeling the distribution of sensor error over multiple states.

3) We devise a game which is representative of the security related interactions. The game has 5 states, each having a distinct action set, resulting in distinct perceived/extended action set corresponding to each distinct information set of states.

4) We present the expression for the generalized Imperfect Information Factor in **Appendix B**.

5) We present the numerical simulations studying the dynamics of such a game when played with malicious users of different natures, and picking the preferred strategy profile.

6) We present the framework to actively interact with suspicious players in honeypot.

7) We present the conditions for interacting with a group of attackers and for collaboration.

## 2. Related Works

A survey of application of game theory for network security based on the targets in a network is at [8]. A survey of different network related attribution to the game theoretic actions with equilibrium calculation is at [9]. A simulation of security game with 5 strategic options of administrator and the adversary is at [6]. A dynamic altering of the properties of a computing platform based on the game-theory to counter the attacks is studied with simulations at [10]. A zero-sum game model is presented at [11]. A game theoretic framework with classification of the attackers based on the skill and motivation, with solution based on solving the Markov chain is at [12]. A game theoretic attempt to secure the infrastructure resources is at [13]. Computation of 2-Player Nash Equili- brium is shown at [14] to be complete in the class PPAD introduced at [15]. An algorithm for general-sum stochastic game in two scenarios is at [16]. Considering behavioral probability weighting of players on their equilibrium strategies in three types of security games is presented in [17]. In this work, the weighting of players results in capturing the situations where different objectives are needed for attacks to succeed, as noted by the *Total Effort*, *Weakest link and Best Shot* games. We here use the different behaviors (as explained in Section 7.1) of players as a parameter which they vary to achieve optimal result against the adversary.

## 3. Game Inspired Security Architecture

Here we present the architecture. We refer to the security architecture as Game Inspired Defense Architecture (GIDA) (**Figure 1**).

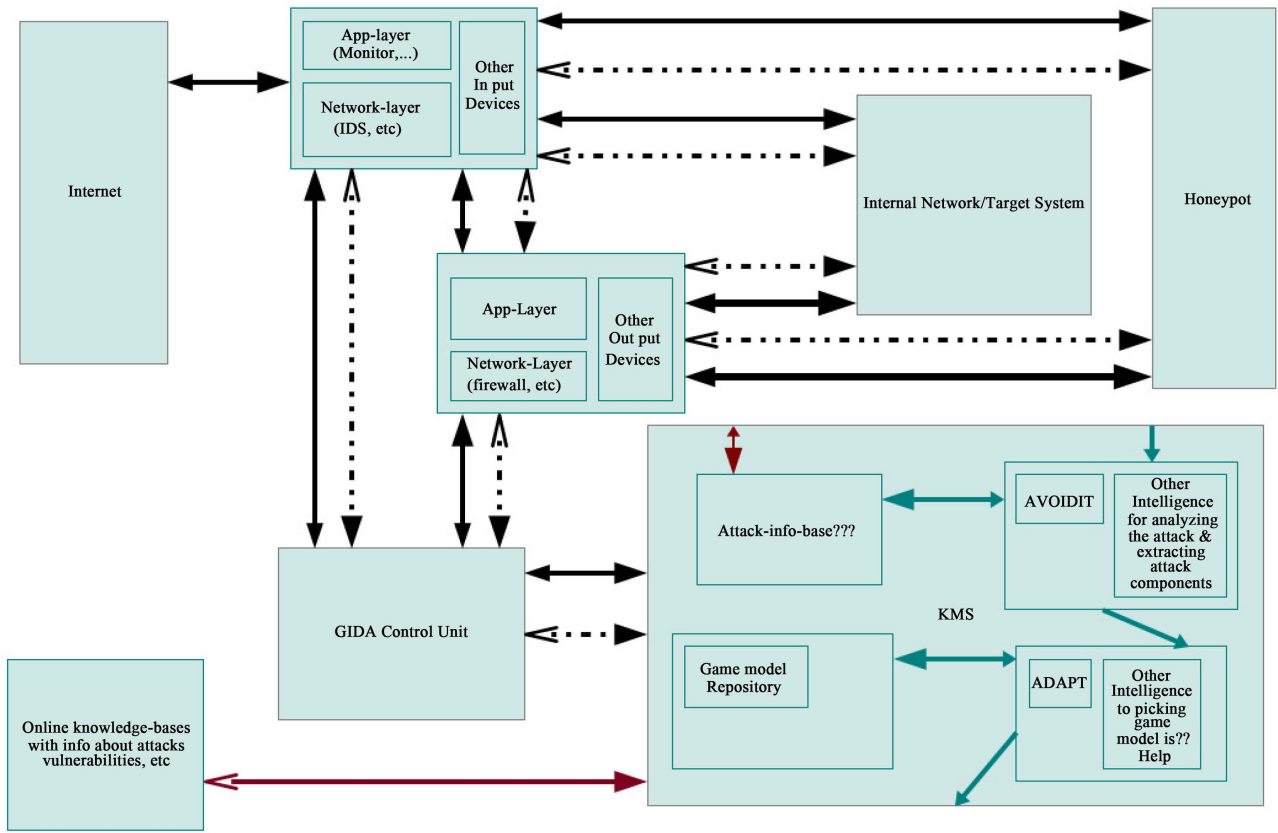

**Figure 1.** GIDA.

- **Target System:** It is the computational space in CBS which is being protected by the GIDA security system. As is prevalent, such a system would be a network of many systems. All the interaction with other blocks happen through the **Sensors** & **Actuators (S-A)** blocks. It is having a two-way data link and a two-way control-signal link with the **(S-A)** block. The target system is the CBS of the organization which would be using GIDA-based as a real time security system to protect itself from attacks and failure. It would typically contain the usual subsystems that make up any of the present day industrial business system, like database server, web server, confidential files, etc.

- **Internet:** This block is the external world with which the target system communicates during its operation. The system has no control over the form and content of what happens there and should only deal with what comes from it, into itself. The interaction of data coming in and going out into the Internet from the system happens through the **(S-A)**.

  Thus there is a data link with the **(S-A)** block.[2]

  The Internet is functionally connected in layers. Typically, the immediate external surface would be that of the Internet Service Provider for the company.

---

[2]Though there is a possibility of getting some control-signal information, like requesting the ISP to provide some traffic related information, it is not represented here with a control-link signal link, as its not a regular event and not quantitatively salient.

Then those provided by the clients and those interacting with in that surface, and so on.

- **Sensors:** This block is the channel through which all the interactions happen between the Internet and the target system. The sensors and actuators are a bit indistinguishable functionally in terms of distinct systems as interactivity of the system components is maturing with technology. For example, even though *Firewall* keeps a gate to disallow some traffic as per the prescribed policies, it also gives a count of what is the attempted traffic and the one that does go through, to the defense system. In the first function it is acting as an actuator while in the second as a sensor. But the predominantly sensing components are put into sensors. An example for sensors operating at application layer is at [18].

- **Actuators:** This block has the actuators.

The anomalies can happen high up at the application layer or the lower layers like Network layers, Transmission layer, Data link layer and so on. With the convergence in the datacom technologies the distinction of these layers is also not clear in the present times. So, what ever the sensors and actuators the operate with a given functionality space is contained in this block.

Similarly, with the maturing of technology of interfaces the same devices/ systems some times do both sensing and actuating functions. So, that classification is also not a hard one. But the systems which are able to sense the anomalies, at what ever the layer are termed as sensors and the systems which are able to act upon to affect the **Target System**'s behavior are considered as actuators.

There is typically some inbuilt set of actions based on the standard prescribed policies, which allow the actuators to take actions. Responses for some obvious anomalies are also covered by them. But when some anomaly is observed which is not obvious then those readings/values/data/related to that event is sent to the **GIDA Control Unit (GCU)**. Then the **GCU** transfers these observations to the **Knowledge Management System (KMS)** and receiving an action plan based on an appropriate *game model* it will direct/send control signals to the actuators to take actions to deal with the scenario. Subsequently the sensor functionality of the systems in this block keep sending any required information back to the **GCU**.

In case of an observed anomaly which has a corresponding prescribed action for the sensor/actuator system to counter, then it shall be acted upon as prescribed. But never the less the event is reported to the **GIDA Control Unit**, at the earliest, for it to be logged in and documented in the **KMS** for further analysis and forensic activities to learn about the system and scenarios.

It is having both a two-way data link and two-way control-signal link with the block **Target System** and a data link with the block **Internet**.

- **KMS:** This is the knowledge center where the information related to the observed anomalies is sent in and a defense action plan is got as the output to the **GCU** block. To do the attack identification effectively, an up-to-date knowledge is needed about the attacks, risks, vulnerabilities etc. There are many online government databases maintaining such knowledge bases. This block keeps itself updated from those knowledge bases and reports to them its own findings.

There is a two-way data link and a two-way control link to the **GCU** block. There is also a two-way data link with the **External Defense knowledge Systems** block.[3] The KMS is the block which takes the input of observations anomalies submitted through the **GCU**. These inputs are analyzed to identify the attack/attacks as per AVOIDIT, and to pick a suitable model as per ADAPT. Then this game model is given as output to the **GCU**. This can be as simple as a onetime action of shutting down a web server or restarting a database or a sequence of actions in time based on the game model.

While identifying attack, to take the advantage of the up to date online resources, the block with information related to attack, vulnerabilities, risk etc. will be communicating with the online resources to get the information about the latest attacks.

- **Honeypot:** This block is the trap to which a suspected user is sent, to understand his intentions when his activity raises alarm through the sensors enough to consider it unsafe to let him interact with the main system.

It has a two-way data link and a two-way control link to the **(S-A)** block. This block mimics the main target system in its constitution *virtually*. Even though does not contain any real confidential information or high value assets, it only poses to have them.

When the anomalous behavior of a user is ambiguous enough to resist the classification as either an attacker or a normal user, further inquiry is needed to learn more about the intention of the user. If it seems unsafe to wait and watch the user to operate on the real system, it is a convenient strategy to lead him to **Honeypot** where he is under the illusion of being in the real system, and operates as his plans. This facilitates GIDA to observe the user's behavior through the same **(S-A)** block, which also is connected to the **Honeypot**. Here its not only observing the actions, an instigating interaction is possible leading to a classification of the user possibly with optimum time and effort. Such a classification would help in the appropriate treatment of the user.

- **External Defense knowledge Systems**: This block is the online knowledge bases related to attacks, vulnerabilities, risks etc. There is a two-way data link with **KMS** block.
- **GCU**: This block as shown in the <span style="color:red">Figure 2</span> is the central control unit, for the whole defense operation of GIDA. It is connected to the **(S-A)** which intimates when there is any anomaly observed. This starts the involvement of this unit. The control agent translates and facilitates the communication between the **Sensors** and the **KMS** in one way, while from **KMS** to the **Actuators** in the other, as shown in the <span style="color:red">Figure 3</span>. As and when required it sends control signals to **(S-A)** to abort a traffic or operation, divert it to **Honeypot** or communicate with the **KMS** to decide what to do.

---

[3]Since this is done periodically and not necessarily during the engagement with an attacker the links are show in red.

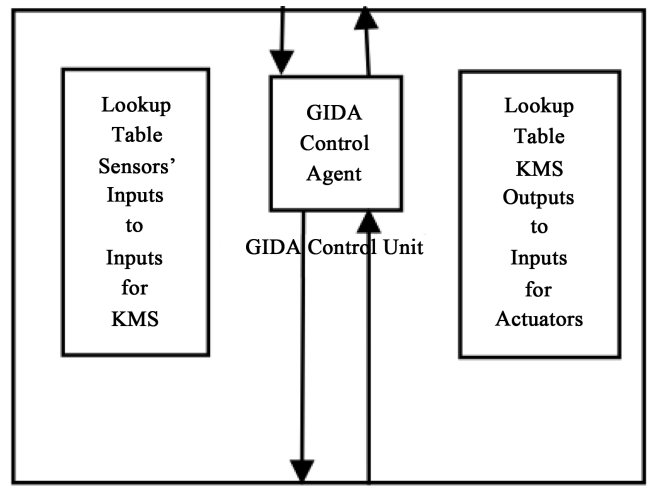

**Figure 2.** GIDA control unit.

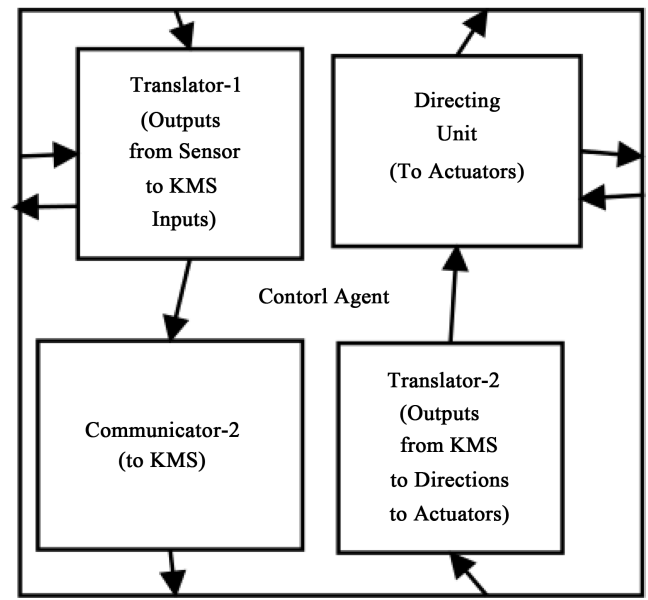

**Figure 3.** GIDA control agent.

This block controls and co-ordinates the defense/security activity of GIDA. It gets its inputs about the observed anomalies from the sensors in the **(S-A)** block and submits it to **KMS** and gets the plan of action as per a *game model* as the counter measure. It then uses the look-up table to send directions for such actions to be performed by the actuators in the **(S-A)** block.

This block consists of mainly three sub blocks. A GIDA-Control-Agent and the two other blocks with look-up tables for translating the two inputs into two outputs. The first one is a look-up table mapping inputs of observation[4] of anomaly/anomalies from the sensors to the inputs to be submitted to **KMS**[5]. The second one is a look-up table mapping the output of the **KMS** which is an ac-tion-plan based on a game model, to the the action commands to be given to the actuators in the **Actuators** block. The control agent box has four constituents as shown below.

[4]These observations collected by the sensors must be useful for the **Attack Identification System** for analysis. So, **GIDA Control Unit** will have to feed it in appropriate way.
[5]With the maturing of the model, the **KMS** can be designed to take the inputs which are as closer to the sensor inputs.

## 3.1. Operational Flow

The typical operational flows are as follows:

1) when there is no attack or any problems in the operation of the target system, the target system is simply interacting with its clients over Internet. Both the interactions and operations are being observed by the systems in the **(S-A)**. The most probable event in GIDA system on such a day would be, a lucky update of the knowledge base in **KMS** related to some attack from an online resource.

2) When there is an anomaly observed by the sensor, which is obvious in nature, and already has a prescribed action to be taken by a co-operating actuator, then such an action would be taken immediately and the event is reported at the earliest to the GIDA Control Unit. The information received like this will be submitted to **KMS** for documentation and forensic purposes.

3) When anomaly is observed by some sensors, and the observed pattern does not have any obvious (reflex) action plan for the sensors/actuators, they report those observations to the **GCU**. Then these inputs and submitted[6] to the **KMS** appropriately by the **GCU**[7]. The information is analyzed and attack/attacks is/are identified using attack identification system. That identification will be used to pick the appropriate game model as per ADAPT. That game model will be given as the output of **KMS** to the **GCU**. Then **GCU** takes prescribed sequence of action as per that game model[8].

Once an observed anomalous behavior is not obvious enough for the actuator/sensor systems to act upon itself, the anomaly is referred to the **GCU**. Then through **KMS**, every response is formulated as a game. As everywhere in game theory literature, from the trivial interaction to the most sophisticated one is modeled as a game, we model all the responses unequivocally as a game.

4) When the information from the observed anomaly is sufficient to classify the user to be not safe to operate in the main system, while it is insufficient to classify if he is really malicious and identify his intentions, he would be directed to honeypot and the step two or three is followed as if he is on the main system.

## 3.2. Game Model Repository

The success of a security system based on game theoretic ideas lies on appropriateness of the game model to represent the current security situation and the effectiveness of the prescribed strategies. Not only different attacks warrant for different games to be played with different appropriate game models, different scenarios with same kind of attack may need different game models with different considerations of rewards and transition functions. But to have a generic game model which can be used to derive such game models is crucial, as that gives the general structure with proven boundary conditions. The Figure 4 shows this connection.

---

[6]The **GIDA Action Agent** uses the first lookup table for this.

[7]A typical example for this would be when two different sensors detect slight anomalies, but individual neither large enough to warrant a standard reflex action, as per sensor/actuator policies.

[8]The **GIDA Action Agent** uses the second lookup table for this.

Functionally GIDA is designed as an event-driven system as show in Figure 5. Computationally, it is an implementation of a *Buchi Automaton*.[9] An overview of the event flow can be summarized as in this figure. A set of anomalies is detected in the operation of system to be protected (during operations by a user/ some users). This triggers the operation of GIDA. Then this information is used to identify the possible attack/attacks going on using the taxonomy AVOIDIT. This gives the attack components, the relevant game models. Using both of these in ADAPT, the relevant game model is obtained.

This Figure 6 gives a succinct event flow diagram of the events in terms of interactions between a user and an administrator. By this figure and the previous one, we can see that though we need different game models with different characteristics, there seems a commonality in the security games and the modeling requirements. Until now most of the effort is focused on using game theory to address some particular attacks. But to systematize this as a paradigm for security solution, a generic game model becomes a necessity.

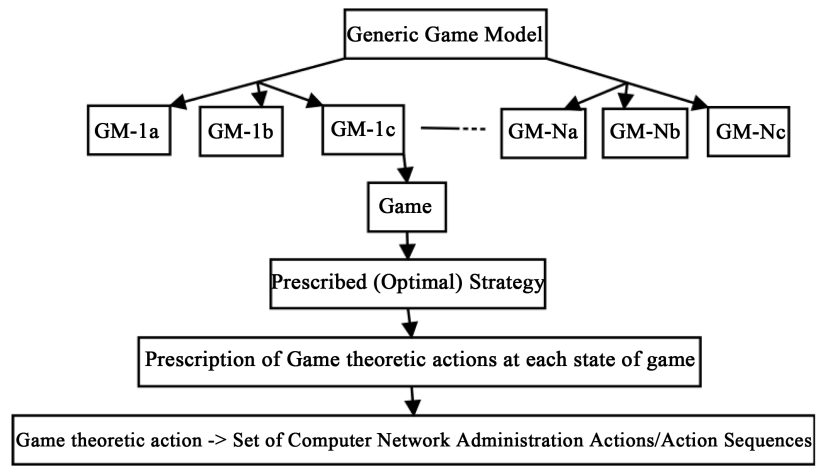

**Figure 4.** Generic game model's utility.

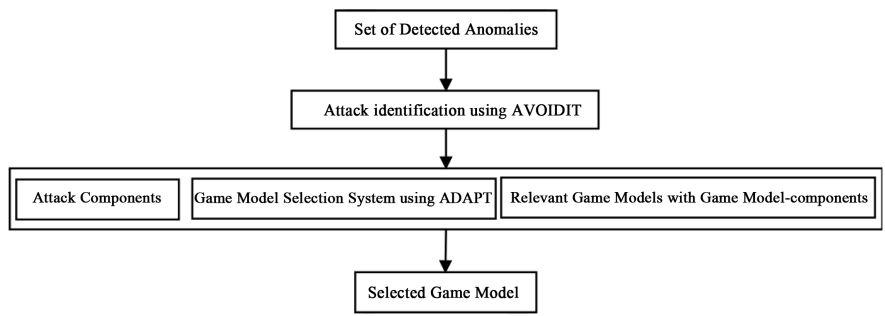

**Figure 5.** Internal flow diagram.

---

[9]The system will have accepting conditions and not an accept state like the general NFA, and is facilitated to run continuously.

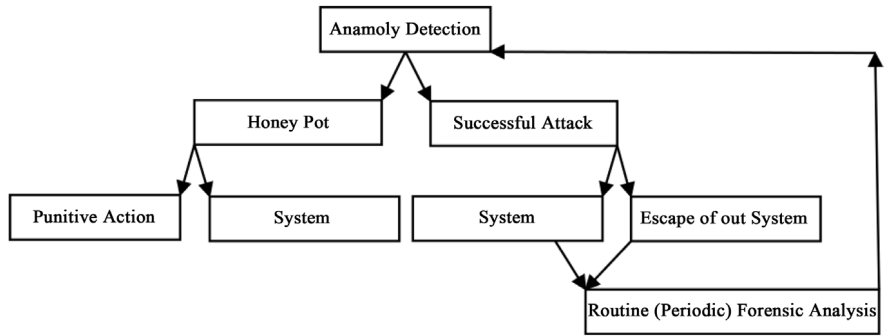

**Figure 6.** Security event flow.

## 4. Game Model

The model considers that a player[10] $k$ observes the game's true state using an imperfect sensor/a set of imperfect sensors. That means, player $k$ can view the present state $\xi_j$ to be any state in the information set $I_{\xi_j}^k = \left\{ \xi_{j_1}, \xi_{j_2}, \cdots, \xi_{j_p} \right\}$ with $\xi_j$ being an element of $I_{\xi_j}^k$. The perceived action set at this state may be expanded, *i.e.*, player may decide to take an action which is allowed at $\xi_{j_i} \neq \xi_j$ where $\xi_{j_i}$ belongs to $I_{\xi_j}^k$. When the true state is $\xi_j$, let the player $k$'s extended action set $B_{\xi_j}^k = \bigcup_{\xi_j \in I_{\xi_j}^k} A_{\xi_j}^k$ where $A_{\xi_j}^k$ denotes the allowed action set of player $k$ at state is $\xi_j$. If the player $k$ takes an action $\alpha^k \in B_{\xi_j}$, when the true state is $\xi_j$ but $\alpha^k$ is not in $A_j^k$, then in terms of the influence on state transition probabilities, $\alpha^k$ is considered equivalent to player $k$ taking no action at state $\xi_j$. However, its influence on player $k$'s payoff $\alpha^k$ may not be equivalent to player $k$ taking no action at state $\xi_j$ depending upon the cost of the attempted execution of $\alpha^k$. Formally, the model is represented by a tuple, $\left( S, E^1, E^2, A^1, A^2, Q, R^1, R^2, \beta \right)$ whose elements are defined below.

1) $S = \left\{ \xi_1, \xi_2, \cdots, \xi_N \right\}$ is the set of states.

2) $E^k = \left\{ E_{\xi_1}^k, E_{\xi_2}^k, \cdots, E_{\xi_N}^k \right\}$, $k = 0,1$ where the $j^{th}$, $0 < j < N$, set $E_{\xi_j}$ with $E_{\xi_j}^k = \left\{ p_{j_i}^k \,\middle|\, 1 \leq i \leq m_j, \sum_{i=1}^{m_j} p_{j_i}^k = 1, p_{j_i}^k > 0 \right\}$, represents the error probabilities of $k^{th}$ player's sensor at the true state $\xi_j$ over the corresponding information set, $I_{\xi_j}^k$. $I^k = \left\{ I_{\xi_1}^k, I_{\xi_2}^k, \cdots, I_{\xi_N}^k \right\}$, $k = 0,1$ where the $I_j^k$ represents the information set of player $k$ when the true state is $\xi_j$, *i.e.*,

$$I_{\xi_j}^k = \left\{ \xi_{j_1}, \xi_{j_2}, \cdots, \xi_{j_i}, \cdots, \xi_{j_{m_j}} \right\}$$

where $m_j = \left| I_{\xi_j}^k \right|$, $\xi_{j_i} \in S$, with $m_j \leq N$ being an integer indicating the number of states that have a possibility of being considered the current state at state $\xi_j$ with the condition that $\xi_j \in I_{\xi_j}$.

3) $A^k = \left\{ A_{\xi_1}^k, A_{\xi_2}^k, \cdots, A_{\xi_N}^k \right\}$, $k = 0,1$ is the action set of player $k$, where $A_{\xi_j}^k = \left\{ \alpha_{j_1}^k, \alpha_{j_1}^k, \cdots, \alpha_{j_{M^k}}^k \right\}$ is the action set of player $k$ at state $\xi_j$. Let

---

[10] $k = 0,1,\cdots,K$ for one administrator and users respectively.

$$B^k = \left\{ B_{\xi_1}^k, B_{\xi_2}^k, \cdots, B_{\xi_n}^k \right\}, k = 0,1$$

where $B_{\xi_j}^k$ represents the action set of player $k$ at $I_{\xi_j}^k$. That means $B_{\xi_j}^k = \bigcup_{\xi_j \in I_{\xi_j}^k} A_{\xi_j}^k$. By introducing different action sets at each state we may get distinct $B_{\xi_j}^k$ at for each distinct $I_{\xi_j}^k$. Let $T_{\xi_j}^k = \left| B_{\xi_j}^k \right|$.

4) The state transition probabilities are represented by function $Q : S \times B^1 \times B^2 \times S \rightarrow \begin{bmatrix} 0 & 1 \end{bmatrix}$ which maps a pair of states and a pair of actions to a real number between 0 and 1. The model assumes that for any state $\xi_j^k$ if the player $k$ takes an action $\alpha_j^k \in B_{\xi_j}^k$, which is does not belong to $A_{\xi_j}^k$, then

$$Q\left( \xi_{j_1}, \alpha_{i_1}^k, \alpha_{i_2}^l, \xi_{j_2} \right) = Q\left( \xi_{j_1}, \text{Normal\_operation}, \alpha_{i_2}^l, \xi_{j_2} \right)$$

where $l$ represents the other player.

5) The reward[11] of the player $k$ is determined by the function $R^k : S \times B^1 \times B^2 \rightarrow R$ which maps a state and a pair of actions to a real number.

6) $\beta, 0 < \beta < 1$ is the discount factor for discounting the future rewards in this infinite horizon game.

## 4.1. The Error Distribution

As $E^k$ represents the set of error probabilities of the player $k$, let us consider the set of error probabilities $E_{\xi_j}^k$ with the current state being $\xi_j$. Let the error of the sensor for player $k$ at the state $\xi_j$ is $\gamma_j^k$, $0 \le \gamma_j^k < 1$. The error $\gamma_j^k$ is always less than 1 because the real state $\xi_j$ is always taken as an element of the information set $I_{\xi_j}$ at $\xi_j$. Then at the current state $\xi_j$, let the probabilities with which administrator perceives the current state to be $\xi_1$, $\xi_2$, $\xi_3$ and $\xi_4$ be $p_{j_1}^k$, $p_{j_2}^k$, $p_{j_3}^k$, and $p_{j_4}^k$ respectively. Then the error at state $\xi_j$ is

$$\gamma_j^k = \left( \sum_{i=1}^N p_{j_i}^k \right) - p_{j_j}^k = 1 - p_{j_j}^k. \text{ For } 1 \le i, j \le N, \text{ let } \omega_{j_i}^k \text{ be the set of sensor in-}$$

puts to the player $k$ indicating the current state to be $\xi_i$, while the real current state is $\xi_j$. In practice the sensor can be a device or a collection of devices which collects values of some parameters of the system. All such parameters can be considered to form an orthogonal basis of the vector space, where some closed volume is taken to be associated with each state. All of those points in that closed region get mapped to one state. All of them have the values of the parameters which lead the player to perceive the current state to be the particular state.[12]

Let the current real state be $\xi_j$, where $1 \le j \le N$. At this state the sensor inputs at two different instances be, say $\omega_{j_i}^k$ and $\omega_{j_h}^k$, where $1 \le i, h \le N$ due to erroneous sensors. This leads to the perception of the current state to be $\xi_i$ and $\xi_h$ respectively. Depending on the nature of the system, consider some representative statistical measure of central tendencies like Mean, Mode, Median and so on, of $\omega_{j_i}^k$ and $\omega_{j_h}^k$. Let $Ed_{j_{ih}}^k$ to be the Euclidean distance between those measures. In this work, *larger errors are assumed to be less probable than the smaller errors* in sensor operations, that is, $Ed_{j_{ji}}^k > Ed_{j_{jh}}^k \Rightarrow p_{j_i}^k > p_{j_h}^k$. In the following game with 5-states, the sensor error is distributed over the three states other than the real current state. This means for example, if the sensor error at $\xi_1$ is 0.3, then the probability with which the administrator perceives the current state to be not $\xi_1$ is 0.3. This could result in perceiving the current state to be $\xi_2$, $\xi_3$ and $\xi_4$ states respectively with probabilities of 0.15, 0.1 and 0.05. From the sensor relation, we have the sensor error at state $\xi_j \big| i \neq j, \gamma_j^k = \sum_{i=1}^N p_{j_i}^k$ related to the probabilities of virtual states.

---

[11]The expected quantified value of the outcome is called reward.

[12]It is true that some of the parameters may yield discontinuous intervals of values associated with a single state. This makes a set of disconnected regions correspond to a single state. In such cases we have to device more involved mechanisms, which is apt for future enquiries.

The sensor outputs values of parameters which define states. Particular range of values of particular set of parameters will correspond to a state. In fact the general way to define it, is a set of values, covering the range, for each of the parameters to correspond to a state. And if there are some parameters whose values do not affect in deciding a particular state then then range can accommodate any value.

The states of the system is formulated using the following.
- values[13] corresponding to a set of parameters[14]
- a vector space with these parameters constituting the basis[15]

Thus we can formulate probabilities of observed instant being in the defined states as follows.
- $S \rightarrow S - O$

where

$S - O = \left\{ S - O_0, S - O_1, S - O_2, \cdots, S - O_N \right\}$,

$S - O_i = \left\{ s - o_{i_0}, s - o_{i_1}, s - o_{i_2}, \cdots, s - o_{i_{Fi}} \right\}$

$s - o_{i_j} = \left( s - o_{i_j}(0), s - o_{i_j}(1), s - o_{i_j}(2), \cdots, s - o_{i_j}(g) \right)$ is the $j^{th} g$ -dimensional sequence corresponding to $i^{th}$ state, $\xi_i$.

$s - o_{i_j}(l)$ = a MCT, representative value of the expected range of values for the $l^{th}$ parameter in the $j^{th}$ behavior $\in \xi_i$. $l \in \mathbf{N}$, $0 \le l \le g$

$g$ = number of parameters observed by sensors $0 \le j \le Fi$, where $Fi + 1$ disjoint ranges correspond to state $\xi_i$

When there is an anomaly detected, in terms of sensor values, which do not exactly fit into any particular state and there is uncertainty about to which state the current values belong to, imperfect information must be considered.
- **C**urrent observed **V**alues of parameters as anomaly

$CV = \left( cv_0, cv_1, cv_2, \cdots, cv_g \right)$ implies the current values of $g$ parameters
- consider the minimum of Euclidean distances with elements in each state $\left( \Delta_{0_{f0}}, \Delta_{1_{f1}}, \Delta_{2_{f2}}, \cdots, \Delta_{N_{fN}} \right)$. where $fi$ is the $fi^{th}$ element, $0 \le fi \le Fi$ has the minimum of distances between $CV$ and $s - o_{i_{fi}}$ element in the $i^{th}$ state.

$$\Delta_{i_{fi}} = \sqrt{\frac{\sum_{j=0}^{j=g} cv_j - \left( s - o_{i_j} \right)}{g}}_{\min} \tag{1}$$

---

[13] Range of values, represented by a MCT.

[14] (e.g. download bandwidth used, upload bandwidth used, number of ftp requests by a user, …).

[15] (orthogonal/orthonormal) depending on the parameters' co-relation.

The error in perception of states is given by

$$p_{j_i} = \frac{\sum_{j=0}^{j=N} \Delta_{j_{fj}} - \Delta_{i_{fi}}}{\sum_{j=0}^{j=N} \Delta_{j_{fj}}} \qquad (2)$$

And the error distribution is given by

- Then the error at state $\xi_j$ is

$$\gamma_j^k = \left(\sum_{i=1}^{N} p_{j_i}^k\right) - p_{j_j}^k = 1 - p_{j_j}^k \qquad (3)$$

- Larger errors are assumed to be less probable than smaller errors.

  An simple illustration to show its application can be as below.

  Example:

  Let some office has a worker's network traffic as below.[16]

  **Low_Privilege State**

  8AM - 6PM ={Download ≤ 1 Mbps, Upload ≤ 150 kbps}

  This behavior is low-privilege, when worker works at his computer in the office.

  **High_Privilege State**

  8AM - 6PM ={Download ≤ 1 Mbps, Upload ≤ 150 kbps}

  This behavior is high-privilege, when manager works at his computer.

  **Sensor Readings** On some day, at 3:30 PM there is 1.7 Mbps download and 150 kbps upload.

1) $\dfrac{(0.7+0.3)-0.7}{(0.7+0.3)} = 0.3$ Probability in Low_Privilege State. Mean Download: 1 Mbps

2) $\dfrac{(0.7+0.3)-0.3}{(0.7+0.3)} = 0.7$ Probability in High_Privilege State. Mean Download: 2 Mbps

## 4.2. Reward and Transition Functions

In a given state of the game, the action selected by the players decide their rewards and not the resulting state transition that happens depending on such a selection. The rewards are affected by the

- The cost of attempting/executing an action. Let the unit vector representing this be $a_1$.
- The desirability of the possible state change brought by the selection of an action. Let the unit vector representing this be $a_2$.
- The amount of information being divulged to the other players regarding the intention made explicit by the selection of an action. Let the unit vector representing this be $a_3$.

Depending upon the system these aspects may be mutually affecting each other or independent of each other. The reward is located in the vector space thus constructed with these unit vectors as basis. With system having, each of these aspects $a_1$, $a_2$ and $a_3$ with $i, j \in \{1, 2, 3\}$ & $i \neq j$, fully affecting the other, (*colinear*), $a_i \cdot a_j = 1$, being mutually independent (*orthogonal*), $a_i \cdot a_j = 0$ and when partially affect each other will have $0 < a_i \cdot a_j < 1$. This is illustrated at **Appendix A**.

---

[16]For other kinds of parameters, with crucial readings transmitted across channel, edit distance can also be a useful measure to find errors.

The transition depends on the actions selected by the players in the current state and the stochasticity involved. Stochasticity is not arbitrary nor have to do with the instability of the system per se, but provides for the possibility of the state transition propelled by the actions of the players with conflicting interests.

## 4.3. Equilibrium

Given the information set associated with each state $j, 0 \leq j \leq 4$ the corresponding mixed strategy is given as

$$\pi^k \left( \xi_j \right) = \left[ \pi^k \left( \xi_i, \alpha_{i_1}^k \right), \pi^k \left( \xi_i, \alpha_{i_2}^k \right), \cdots, \pi^k \left( \xi_i, \alpha_{i_l}^k \right) \right],$$

where $\xi_i \in I_{\xi_j}^k$, $l = \left| A_{\xi_i}^k \right|$ & $\alpha_i^k \in B_{\xi_j}^k$. Thus, $\left| \pi^k \left( \xi_j \right) \right| = T_{\xi_j}^k$. We are considering general-sum games. When the game starts, say at $\xi_j$, with the input of state $\xi_j$ and each player's selected action $\left( \alpha_j^0, \alpha_j^1, \alpha_j^2, \cdots, \alpha_j^K \right)$, we calculate the reward at state $\xi_j$ of each player, $r_j^k, 0 \leq k < K$ as per the reward function $R^k$. The net reward incurred by player $k$ is $v_{\left( \pi^0, \pi^1, \pi^2, \cdots, \pi^K \right)}^k = \sum_{j=0}^{-\ln \beta^{\gamma}} \beta^j q_j v_{\left( \pi^0, \pi^1, \pi^2, \cdots, \pi^K \right)}^k \left( \xi_j \right)$,

where $q_j$ is the probability of reaching the state $\xi_j \big| \xi_j \in S$ during the game's execution at $j^{th}$ step. $q_j = \prod_{i=0}^{i=j} q_i, q_0 = 1$ as game does start at $\xi_j$ state with $j = 0$, $q_j$ is the probability of game reaching $\xi_j$ from $\xi_{j-1}$. $\gamma$ is the effective discounting factor deciding the threshold at horizon, due to insignificance of the effective value addition of further rewards. Since the game is an imperfect information game, $B_{\xi_1}^0$ gives the perceived action set. Here if the states $\xi_1, \xi_2, \xi_3$ are perceived to be the current state with a probability say, 0.1, 0.3 and 0.6 respectively, we get the probabilities associated with actions of each state multiplied by these corresponding probabilities. With this extended action set, the selection is done, by each players, their rewards are calculated for this step.

We know that the actions belonging to $\xi_2$ & $\xi_3$ cannot be executed and its selection is equated to selecting the normal operation of $\xi_1$, for the effect on state transition. Now based on these probabilities the effective action selection is evaluated for the player. Thus the Nash equilibrium would be the strategy profile

$$\left( \pi_*^0, \pi_*^1, \pi_*^2, \cdots, \pi^K \right) \big| c^0 \& c^1 \& \cdots \& c^k \& \cdots \& c^K$$

where $c^k \Leftrightarrow v_{\left( \pi^0, \pi^1, \pi^2, \cdots, \pi_i^k, \cdots, \pi^K \right)}^k \leq v_{\left( \pi^0, \pi^1, \pi^2, \cdots, \pi_*^k, \cdots, \pi^K \right)}^k$, $\forall i, \forall c^k \big| 0 \leq k \leq K$.

# 5. The Game

In this section we present the game with five states based on the model in Section 4, to address the security situations. In general the user presence in a network system can be classified as one with low and high privileges. The exact privileges could be different depending on the networks and situations, but this distinction provides distinct premises for the access, possible actions and responsibilities of the user. Thus the activities, both malicious and benign, security related or not, could be respectively categorized and represented by formulating two different states. When a malicious user successfully attacks and a suspicious user gets trapped in a honeypot by the administrator for investigation, the resulting two situations are formulated as two other states. When the user logs out of the system he is considered to be out there in the Internet in another state. Since the security oriented activity happens only after the user is at least in the low privilege state, to start with, the game as in 7 explicitly begins at the Low_Privilege_State in the <span style="color:red">Figure 7</span>. Thus these 5 states provide a relevant representation of general conditions of user activity and hence the user-administrator interaction.

$$S = \{ \text{Internet\_State}(\xi_0), \text{Low\_Privilege\_State}(\xi_1), \text{High\_Privilege\_State}(\xi_2),$$
$$\text{Attack\_State}(\xi_3), \text{Trap\_State}(\xi_4) \}$$

$\xi_0$ = **Internet_State**: In this state, the user is out there in the world connected to Internet. He has to go through authentication as in *Sign_In* successfully to get in to the system. For this, the user must take the Sign-In-Request action and the administrator must grant/take-the-action Sign_In to the user. The normal operation in this state by the user means that he is not sending a legitimate request for Sign_In, and just like millions on the net, is just out there at some IP-address. The normal operation in this state by the administrator means that he is not signing-in this user by granting the permission to be in the system. The cracking efforts to break in to the system by a user is also captured by the normal operation of the user. The blocking, refusal to letting in or system just being unavailable is also captured by the administrator's normal operation. This is to emphasize the main focus to be the interaction of a user while he is already in the system.The game reaches this state:

- when the administrator successfully signs out a user from any of the other 4 states, in response to user's any choice of action including a Sign_Out_Request at that state.

 -when the user having completed his legitimate jobs in the system goes away.

 -when the malicious user having completed his malicious tasks escapes out of the system out in to the Internet.

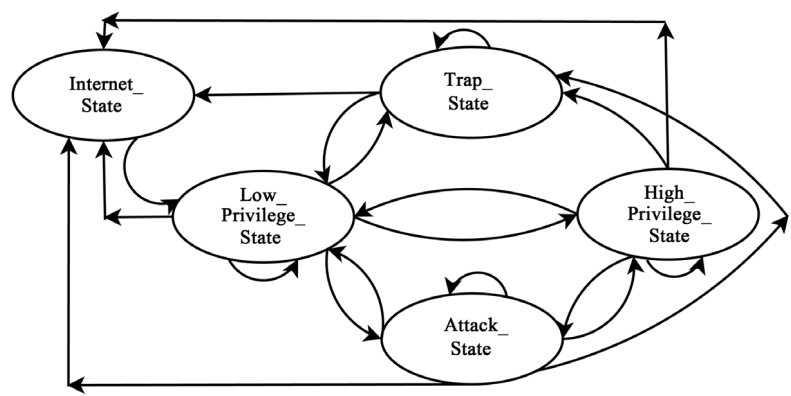

**Figure 7.** The game: States & transition.

- when a person connected to the Internet tries to sign in to the system with a Sign_In_Request but the administrator refuses to let in.
- when a person connected to the Internet tries to sign in to the system with a Sign_In_Request and the administrator allows to let in but the operation fails (due to bad connection, system problem, etc. captured by the stochasticity) and the user remains outside the system on Internet.

Thus the action set of both the administrator & the user is as below.

$$A_{\xi_0}^0 = \left\{ \text{Normal\_Operation\_IS\_U}\left(\alpha_{0_0}^0\right), \text{Sign-In\_IS}\left(\alpha_{0_1}^0\right) \right\}$$

$$A_{\xi_0}^1 = \left\{ \text{Normal\_Operation\_IS\_A}\left(\alpha_{0_0}^1\right), \text{Sign-In\_Request\_IS}\left(\alpha_{0_1}^1\right) \right\}$$

$\xi_1$ = **Low_Privilege_State**: In this state the user is in the system passing through the initial authentication. This is the state when the administrator can take notice of the user's activity. Any anomaly in the user behavior will start attracting administrator's attention/concern. The game can reach this state,

- when the user is already in the system after logging in going through the initial authentication successfully. The administrator can observe the user behavior, though he is not particularly concerned, to begin with.
- when the transition from $\xi_1$ to $\xi_2$ does not succeed, either due to the given the transition probability (system's stochasticity), or the administrator may decide not to promote the user to $\xi_2$.
- when the user in $\xi_2$ acts to return to low privilege, the administrator demotes and the transition succeeds.
- when the administrator decides that a user does not have any malicious intention at $\xi_4$ & lets him back into the system.
- when the user who successfully attacked the system at $\xi_3$, returns to legitimate & low key behavior with in the system either to continue to be so or to elope later. This happens when the user takes the action Return_LPS and the administrator takes the action Normal_Operation_AS_A$\left(\alpha_{1_1}^2\right)$ the transition succeeds as per the occurrence ratio.

$$A_{\xi_1}^0 = \left\{ \text{Normal\_Operation\_LPS\_A}\left(\alpha_{1_0}^0\right), \text{Sign-Out\_LPS}\left(\alpha_{1_1}^0\right), \right.$$
$$\left. \text{Promote}\left(\alpha_{1_2}^0\right), \text{Defend\_LPS}\left(\alpha_{1_3}^0\right), \text{Trap\_LPS}\left(\alpha_{1_4}^0\right) \right\}$$

$$A_{\xi_1}^1 = \left\{ \text{Normal\_Operation\_LPS\_U}\left(\alpha_{1_0}^1\right), \text{Sign-out\_Request\_LPS}\left(\alpha_{1_1}^1\right), \right. {}_{17}$$
$$\left. \text{Privilege\_Request}\left(\alpha_{1_2}^1\right), \text{Attack\_LPS}\left(\alpha_{1_3}^1\right), \text{Resist\_LPS}\left(\alpha_{1_4}^1\right) \right\}$$

$\xi_2$ = **High_Privilege_State**: In this state the user has gained more privileges than in the $\xi_1$. For example, when a user who entered to check his bank account balance has gone into (the higher privileges) editing mode of account information, like the passcode, usually has to go through more authentication. The game reaches this state

---

[17]Here we note that the Normal_Operation is an abstraction of the actions that are typical to that state with no security consequences. Since they are going to be different at different states for both the players, we denote each with different notation to distinguish them.

- when the user successfully receives the privileges by the administrator to move from the Low_Privilege_State.The user should chose the action Privilege_Request and administrator should choose the action Promote, and depending on the game's stochasticity, if the transition succeeds, $\xi_2$ is reached.
- when the user continues to operate with the privileges granted by the administrator with neither attacking nor being trapped by administrator. That is, when both the players are choosing Normal_Operation, depending on the occurrence ratio the game could remain here.[18]
- when the user after successfully attacking the system in $\xi_3$ and completing the intended damage returns to operating as a High_Privilege user. The user should choose Return_HPS, the administrator Normal_Operation_AS_A and the transition should succeed.

$$A_{\xi_2}^0 = \left\{ \text{Normal\_Operation\_AS\_U}\left(\alpha_{2_0}^0\right), \text{Sign-Out\_HPS}\left(\alpha_{2_1}^0\right), \right.$$
$$\left. \text{Demote}\left(\alpha_{2_2}^0\right), \text{Defend\_HPS}\left(\alpha_{2_3}^0\right), \text{Trap\_HPS}\left(\alpha_{2_4}^0\right) \right\}$$

$$A_{\xi_2}^1 = \left\{ \text{Normal\_Operation\_AS\_A}\left(\alpha_{2_0}^1\right), \text{Sign-Out\_Request\_HPS}\left(\alpha_{2_1}^1\right), \right.$$
$$\left. \text{Privilege\_Relinquish}\left(\alpha_{2_2}^1\right), \text{Attack\_HPS}\left(\alpha_{2_3}^1\right), \text{Resist\_HPS}\left(\alpha_{2_4}^1\right) \right\}$$

$\xi_3$ = **Attack_State**: In this state the user has already attacked the system successfully. He is either accessing other's confidential data or damaging the system or doing some other harm. Here either he is doing some normal operations which is actually some attacking activity or eloping along with the confidential data, or returning to legitimate behavior with either high or low privileges into $\xi_1$ & $\xi_2$. He would be concerned to wipe out as much as possible any trace of his attack before he goes away. In this state administrator can revive the system, which is like restarting a file server for example, if the file server is found compromised. This can accompany with sending the suspected user either out of the system or to honeypot taking the game to $\xi_4$ with the action Trap_AS. The game reaches this state

- when the user successfully attacks from the $\xi_1$.
- when the user successfully attacks from the $\xi_2$.
- when the user succeeds to behave inconspicuously with Normal_Operation continuing to damage the system.[19]
- when the transitions to $\xi_1, \xi_2$ & $\xi_3$ fails as per the stochasticity of the system.

$$A_{\xi_3}^0 = \left\{ \text{Normal\_Operation\_AS\_A}\left(\alpha_{3_0}^0\right), \text{Sign-Out\_AS}\left(\alpha_{3_1}^0\right), \right.$$
$$\left. \text{Revive\_LPS}\left(\alpha_{3_2}^0\right), \text{Revive\_HPS}\left(\alpha_{3_3}^0\right), \text{Trap\_AS}\left(\alpha_{3_4}^0\right) \right\}$$

---

[18]Alternatively, when the transitions from $\xi_2$ fail to reach the three other states, the game could remain in this state as shown in the **Figure 1**.

[19]The game reaches this state when the user's attempt to transit to the normal state is unsuccessful. This is typical of a situation when the traces are not getting wiped out even though the user is trying to get back to the guise of a normal user.

$$A_{\xi_3}^1 = \left\{ \text{Normal\_Operation\_AS\_U}\left(\alpha_{3_0}^1\right), \text{Sign-Out\_Request\_AS}\left(\alpha_{3_1}^1\right), \right.$$
$$\left. \text{Increase\_Attack}\left(\alpha_{3_2}^1\right), \text{Return\_LPS}\left(\alpha_{3_3}^1\right), \text{Return\_HPS}\left(\alpha_{3_4}^1\right) \right\}$$

$\xi_4$ = **Trap_State**: In this state, the administrator, after suspecting the user while in $\xi_1, \xi_2$ & $\xi_3$ to be an attacker, and has successfully sent him to the honey pot. In this state user could either enact Normal_Operation actions, or with a suspicion of being trapped can do Sign-Out-Request action, which is either trying to close all his operations and logging out at the earliest or engage with the administrator in his ploy behaving like a benign user. User can also confirm his malicious intention by enacting it during the inquiry. The game reaches this state

- when the administrator successfully traps in to Honeypot a user suspecting him of some malicious intention.
- when the user with no malicious intention continues to do Normal_Operation oblivious of being trapped and the administrator continuous to observe him at Honeypot. Alternatively $\xi_4$ is reached when a user with malicious intention becomes suspicious of being trapped starts behaving.
- when the user who originally had malicious intention may try to attack at the false targets provided by the administrator and try to escape by choosing the Sign-Out-Request action, but does not succeed.
- when the administrator identifies the user who has successfully attacked the system and decides to trap him into honeypot while the user is either trying to return to $\xi_1$ & $\xi_2$.

In each of these actions the user engages with the administrator's ploy. In the end administrator may declare either the user innocent and let him back in to the system through the $\xi_1$ or catch him and take further punitive actions. The fact that the user already created suspicion, makes the administrator to let him back to the $\xi_1$, and let him earn his higher privileges, than to directly enter $\xi_2$.

$$A_{\xi_4}^0 = \left\{ \text{Normal\_Operation\_TS\_A}\left(\alpha_{4_0}^0\right), \text{Sign-Out\_TS}\left(\alpha_{4_1}^0\right), \right.$$
$$\left. \text{Test\_TS}\left(\alpha_{4_2}^0\right), \text{Judge\_TS}\left(\alpha_{4_3}^0\right), \text{Allow\_TS}\left(\alpha_{4_4}^0\right) \right\}$$

$$A_{\xi_4}^1 = \left\{ \text{Normal\_Operation\_TS\_U}\left(\alpha_{4_0}^1\right), \text{Sign-Out\_Request\_TS}\left(\alpha_{4_1}^1\right), \right.$$
$$\left. \text{Commit}\left(\alpha_{4_2}^1\right), \text{Behave}\left(\alpha_{4_3}^1\right) \right\}$$

Consequently, the action sets of the players are summarized as below.

User_Action_Space = $A^1$

$$= \left\{ \alpha_{1_1}^1, \alpha_{1_2}^1, \alpha_{1_3}^1, \alpha_{1_4}^1, \alpha_{1_5}^1, \alpha_{2_1}^1, \alpha_{2_2}^1, \alpha_{2_3}^1, \alpha_{2_4}^1, \alpha_{3_1}^1, \alpha_{3_2}^1, \alpha_{3_3}^1, \alpha_{3_4}^1, \alpha_{3_5}^1, \alpha_{4_1}^1, \alpha_{4_2}^1, \alpha_{4_3}^1 \right\}.$$

Administrator_Action_Space = $A^2$

$$= \left\{ \alpha_{1_1}^2, \alpha_{1_2}^2, \alpha_{1_3}^2, \alpha_{1_4}^2, \alpha_{1_5}^2, \alpha_{2_1}^2, \alpha_{2_2}^2, \alpha_{2_3}^2, \alpha_{2_4}^2, \alpha_{2_5}^2, \alpha_{3_1}^2, \alpha_{3_2}^2, \alpha_{3_3}^2, \alpha_{3_4}^2, \alpha_{4_1}^2, \alpha_{4_2}^2, \alpha_{4_3}^2 \right\}.$$

With the above information about the states and actions, the other 6 components of the 9-tuple, $E^1, E^2, Q, R^1, R^2, \beta$ are suitably designed reflecting the system in question. We present a numerical simulation with one such typical set of values in Section 7.

# 6. Game Analysis

As the general case, each of the five states being perceived with a non-zero probability at each state would make the whole action space of the player as the extended action set at each state. But for the game to go into fifth state administrator has to explicitly take the user out of the main system and put into the honeypot. Since neither the stochasticity of the game nor the user can accidentally take the game to this state, administrator wont mistake to be in it while being in the other states. Thus, $p_{ji}^0 = 0$ for $j \neq i$ & $j = 4$, while $0 \leq p_{ji}^0$ & $\sum_{i=1}^{N} p_{ji}^0 = 1, 1 \leq j, i, \leq N$. Thus, $I_{\xi_j}^0 = \{\xi_1, \xi_2, \xi_3\}, 1 \leq j \leq 3$. Hence, the extended action set corresponding to the information set would be for $j \neq 4$,

$$B_{\xi_j}^0 = \left\{\alpha_{0_0}^0, \alpha_{0_1}^0, \alpha_{1_0}^0, \alpha_{1_1}^0, \alpha_{1_2}^0, \alpha_{1_3}^0, \alpha_{1_4}^0, \alpha_{2_0}^0, \alpha_{2_1}^0, \alpha_{2_2}^0, \alpha_{2_3}^0, \alpha_{2_4}^0, \alpha_{3_0}^0, \alpha_{3_1}^0, \alpha_{3_2}^0, \alpha_{3_3}^0, \alpha_{3_4}^0, \alpha_{4_0}^0, \alpha_{4_1}^0, \alpha_{4_2}^0, \alpha_{4_3}^0, \alpha_{4_4}^0\right\}$$

In the Trap state the administrator's response tries to delude the user about his current state to be his preferred state. We consider the user's sensor to be perfect in general, which is the worst case scenario. Thus user's view of the states can be defined as $p_{ji}^1 = 0$ if $j \neq i, j \neq 4$. $p_{ji}^1 = 1$ if $j = i$ & $j \neq 4$. Since the user can also have a honey pot detection system $p_{4_4}^1 > 0$. The administrator's ploy at $\xi_4$ need not be only passive reactive one, a motivation for a proactive one is suggested with the carrot and stick approach in [5]. [19] & [20] discuss such scenarios where the inquiry into the nature and motivation of the attacker are discussed.

Based on the above strictures, the game was simulated and evaluated for rewards to find the preferred strategy profiles.

# 7. Simulation

The game starts only after a sensor reporting an anomalous action by a user. Thus there is a suspicion in the admin and a plausibility of malice in the user. At this stage it is both an incomplete and imperfect information game, with the admin not being sure of both the user's nature and the current state of the game. The rewards for each affecting feature of the game are symbolically distributed over the discrete space defined by $-1000, -100, -10, 0, 10, 100, 1000$. The logarithmically varied values are chosen to represent how the expense of actions and potential outcomes of actions differ. Though it has to be fine tuned for the different attacks based on the specifics this setup gives us a general idea. Taking a Normal_Operation action and for the administrator and the user is a low expense action to attempt and execute in the context of the game. Thus it leads to 0 pay off for both of them in $\xi_1$ & $\xi_2$. Similarly the sign-out request from the user and the sign-out from the admin also leads to a high probability exit of the user from the system. This does not progress the interaction further, but leads to a logical non-antagonizing move. That way the payoff to both is 10, which is not nothing, but not much.

When the user chooses to attack and the admin chooses the action *defend*, the attack is least likely to succeed. This also reveals the user's intention to attack to the admin. Thus this leads to $-100$ for the user and 100 for the admin. Similarly, when admin chooses the action "trap" and the user chooses "resist", the admin's action is least likely to succeed. This also makes the user aware that the admin is trying to trap him, and thus he may start behaving leading to no chance of discovery by the administrator of his real motive with malice, if there was one. This would hence result in −100 to the admin and 100 to the user. These figures are also considered due to the expense of resources to take these actions and the actions impact on the game.

The aggressive actions are the most expensive actions for the players. The exact defensive actions will result in least likelihood of succeeding of the aggressive actions by the opponents. These defensive actions are the second most expensive actions to attempt/execute. The aggression will also yield higher payoff to the player than the defending one, if succeeds. When one player takes an aggressive action, the other player taking *Normal_Operation* or general state transiting actions will lead to greater loss to him and higher gain to the aggressor. This game dynamics at state $\xi_1$ is explained in **Appendix A**.

For computational reasons we restricted the simulation with some constraints. The game was programmed in C and run on a Debian7.6 of GNU/Linux on an Intel CoreTM i5 Processor with 4 GB RAM machine.

1) We ran the simulation with 100 strategies each varying in a gradation of aggression.

2) We find an approximate Equilibrium as explained at **Appendix C**.

3) We ran it for three types of users/attackers and delineate the findings.We simulated the games with each player 100 strategies, playing each 100 times. For better visualization purpose we plot rewards for 10 strategies (whose results are representative of the results with larger range) to cover the aggression, with steps of 0.1 from 0 to 0.9 in **Figures 8-10**.[20]

The simulation was conducted with three user profiles in terms of skill, sophistication and effectiveness as Below_Average, Average and Above_Average users. When in Low_Privilege_State, the Average user would attack and the Administrator defends, playing the perfect antagonizing, there is 0.5 probability that attack succeeds. This way he is on par with the administrator in terms of imposing his will on the system operation. Numerically the Below_Average user would have <0.5 probability to be succeed while the Above_Average user >0.5 in such a situation. As this example illustrates the essential characteristics, all other transitions are similarly poised.

## 7.1. Results

Since the game is stochastic, for the same strategy profile there could be different rewards for the players. So, we ran the game for each strategy profile for 100 times, and picked the average reward. We evaluate a preferred strategy in terms of the discretized gradation of aggression as discussed in **Appendix C**. **Figures 8-10** show the rewards of administrator and the users and the spread of rewards during multiple executions for a typical strategy profile.

[20]TL ⇒ Top Left, TR ⇒ Top Right, BL ⇒ Bottom Left, BR ⇒ Bottom Right, AR ⇒ Admin Reward, UR ⇒ User Reward, ARS ⇒ Admin Reward Values Over 100 Executions with (0.5) mid level aggression for both in the Strategy Profile, URS ⇒ User Reward Values Over 100 Executions with (0.5) mid level aggression for both in the Strategy Profile.

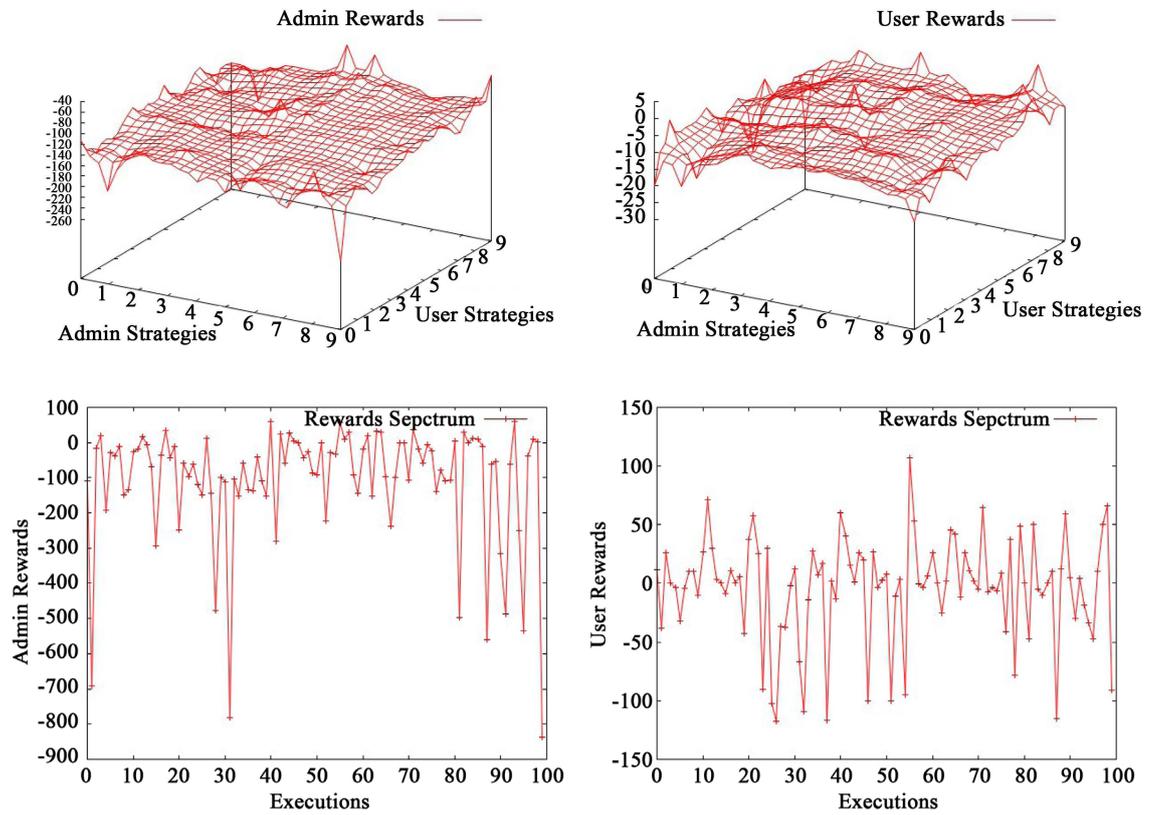

**Figure 8.** Below-average-user: TL-AR, TR-UR, BL-ARS, BR-URS.

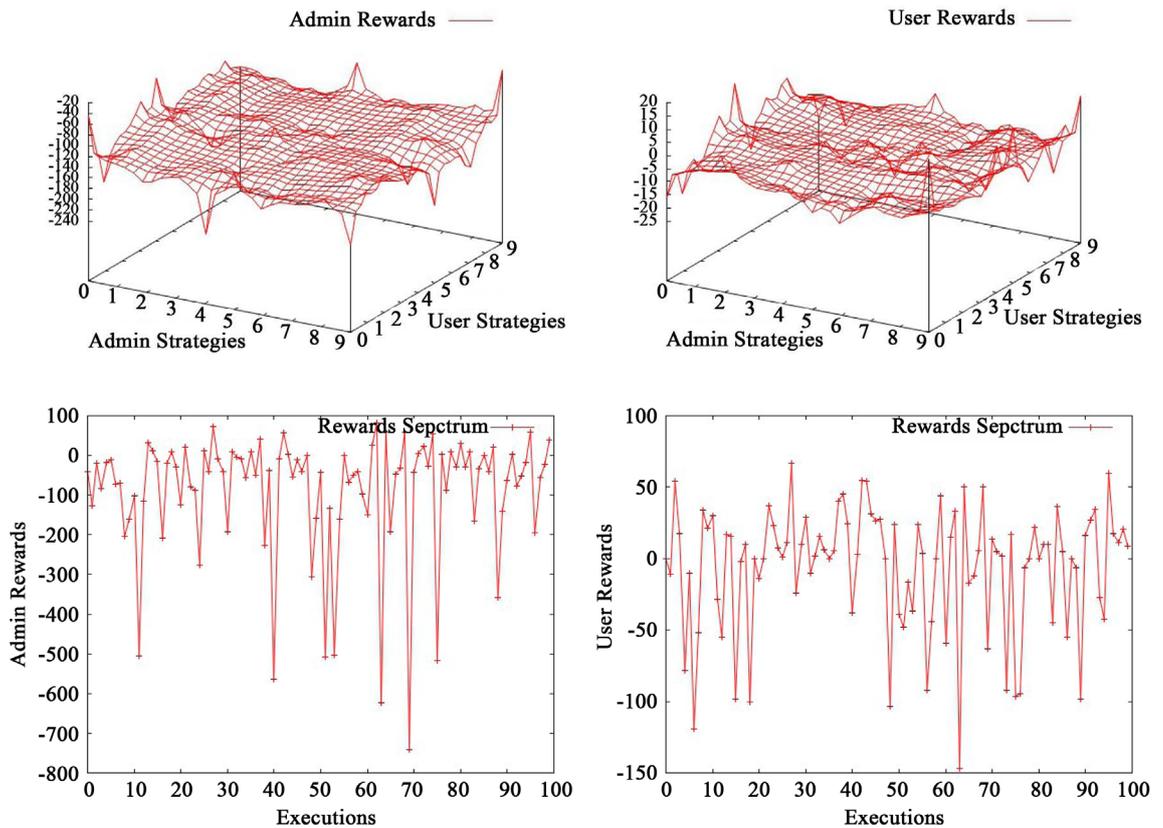

**Figure 9.** Average-user: TL-AR, TR-UR, BL-ARS, BR-URS.

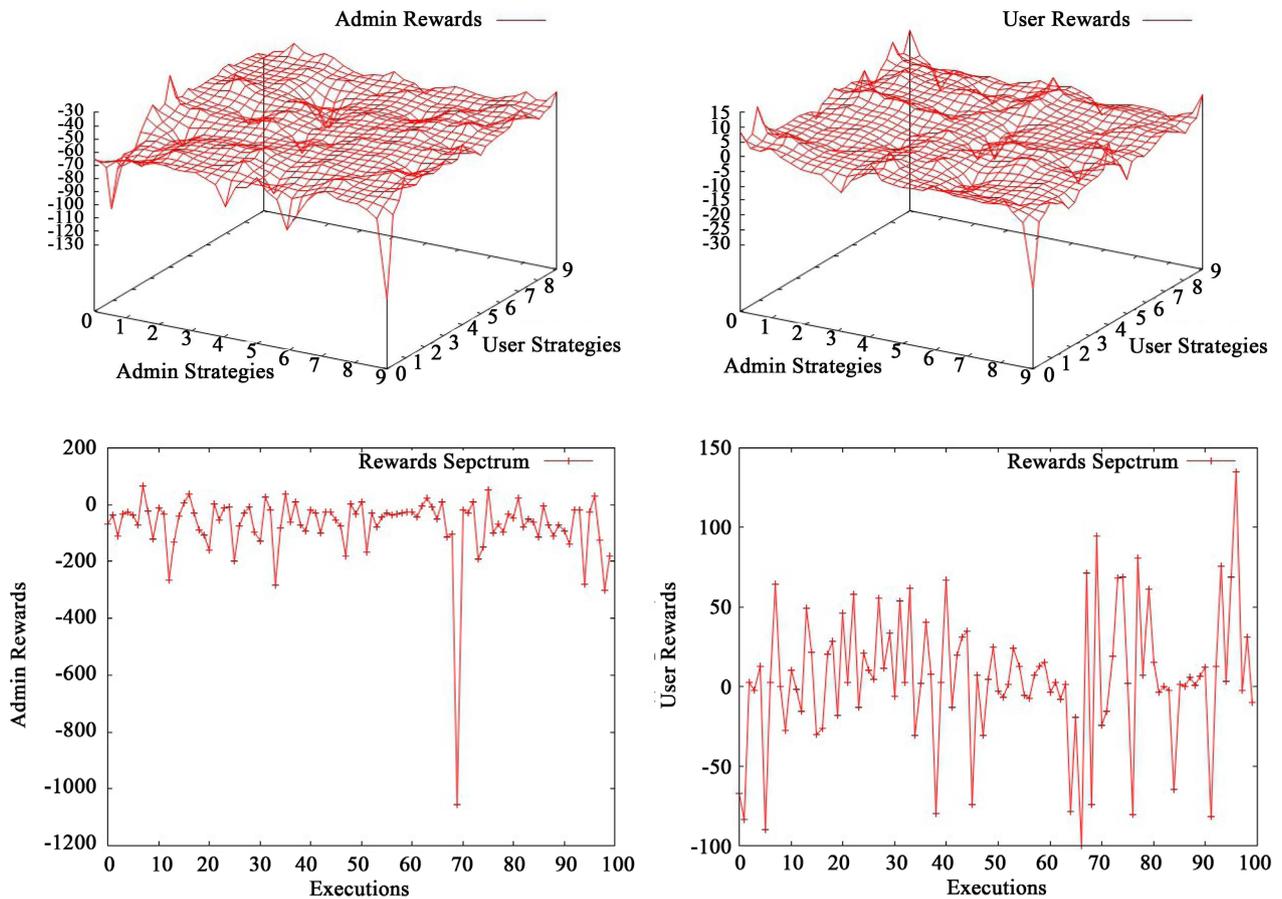

**Figure 10.** Above-Average-User: TL-AR, TR-UR, BL-ARS, BR-URS.

We found many $\epsilon$-NOSPs with $\epsilon = 0.01$ and a different prescribed one in each case. The prescription is based on the least motivation for the attacker to deviate, hence the chance of game going as expected his higher.

### 7.1.1. Below_Average User (BAU)

$(7,2)$ is the prescribed $\epsilon$-NOSP. It recommends for the user and administrator to be least aggressive. The game starts with some anomaly by user and thus if the user is perceived as a BAU, then both players are better off to curtail their aggression.

### 7.1.2. Average User (AU)

$(8,6)$ is the prescribed $\epsilon$-NOSP. It recommends for the user to be less aggressive as he is more affective while the administrator to be quite aggressive.

### 7.1.3. Above_Average User (AAU)

$(4,1)$ is the prescribed $\epsilon$-NOSP. It recommends for the user and administrator to be least aggressive.

Among the possible strategies of players, graded on aggression, this gives a preference for achieving optimal reward. Based on these results, we can infer that during the interaction, due the combination of factors like capacity for domination and secrecy of intent, the players are better off to do most action when its most advantageous to them.

*Conjecture 1: The aggression needed by the administrator for achieving equilibrium follows normal distribution over the power of the user, with the mean at the power of both being equal, while interacting with same reward functions and discount factor.*

*Conjecture 2: The aggression needed by the administrator for achieving the equilibrium would vary from linear to Normal distribution as the depth of the game tree increases, for the same reward functions and the user. The depth of the game tree is a function of discount factor and threshold discount.*

## 8. n-Player Game

The game model presented in Section 4, considers $n, n \in \mathbf{Z}$, $0 < n < \infty$ players where the interactions between many users and administrators can be modeled. As practicality alludes, considering a single administrator would be convenient and useful for most cases. Thus the interaction between many users and an administrator is interesting to investigate. In such a situation, the users can be disparate and working on their disjoint motivations. They could also be working as a group. On the other hand, the administrator can be interacting separately with many users or interact with a group of users with actions collectively affecting all of them. In the light of the game model and game we presented in Sections 4 and 5 respectively, we derive the conditions of the group behavior and reaction of administrator to a group to be optimal.

For each state $j$, let $r_{j_{max}}^k$ be the reward associated with the preferred action choice of the user $k$. During the course of the game, if $\sum_{j=\xi_i}^{j=\xi_f} r_{j_{max}}^k < \sum_{j=\xi_i}^{j=\xi_f} r_{j_{co}}^k$, $k \in S_k$ with $S_k = \{k_1, k_2, \cdots, k_a\}$ where $r_{j_{co}}^k$ is the reward of $k^{th}$ player at $j^{th}$ state when taking co-operative actions, $a$ is the number of users who have motivations that can be pursued in co-operation. There are two trees of game progress that are considered here. In the first case, when each player takes action to fully maximize his total reward during the sequence of game play, he is trying to maximize the probability of those state which maximizes his reward. While doing so two such players may make the game progression such that the final resulting sequence of game play gives both a total reward which is far below their maximum possible value. But this value can be lower than a compromised game transition sequence in which both have higher probability to make higher reward. Recognizing this point, would be the threshold to co-operate.

Secondly, when there are two or more similarly engaged users, responding individually can be expensive for the administrator. Also a response to one user may require an action whose effects on the system conflict with the interaction with other user. The motivating example is when a user who is accessing a particular web server is behaving suspiciously, the admin may want to restart the web server and disconnect the suspicious user. Simultaneously if another user is acting suspiciously with a database server, which is also running on the same machine which has no other processes active at this time, depending on the extent of threat by both the users, it may be optimal for the administrator to re-boot the operating system itself to address both users. The condition is

$$\sum_{g=1}^{g=n} r_{j_{\max}} < \sum_{g=1}^{g=n} r_{j_{\text{common}}}$$

where $r_{j_{\max}}$ and $r_{j_{\text{common}}}$ are the rewards associated with the actions taken for each individual game towards optimal reward, played with different users and the actions taken to collectively/commonly affect all the users concerned. At this condition the administrator is better off playing a single game considering all the users as a collective adversary.

In particular, when many players are involved, the decision process becomes decentralized. If this is true of the defending side, then taking optimal actions such that overall security is served best becomes a challenge. This problem was addressed in [21] and was shown that the individual optima for each defender in his own game to safe guard his own asset may not serve the global good when the assets defended by different defenders are interconnected and interdependent.

# 9. On Engaging Suspicious Users

The only way the game enters the Trapped State $\xi_4$ is from the other states. In the other states, the administrator could become suspicious about a particular user. After sufficient observation of such a behavior of the user, the administrator could become sure that it is not safe to have that user operating freely in the network. For that particular reason administrator sends him to honey pot. Then the game enters the Trapped State $\xi_4$. Then the administrator will have a belief about the user's motive based on these actions of user at one or more of the other states ($\xi_1, \xi_2, \xi_3$), when the game enters $\xi_4$. For example, in a site offering the service of audio streaming but no download, trying to download (successfully or not) could be a reason to put him in Trapped State. The administrator holds the belief that the user wants to steal the audio files. The administrator does not know conclusively yet, either the true nature or motivation of the user. By nature, the user may be not an attacker, a naive attacker or an advanced attacker. The motivation of the user is what exactly he/she intends to do in the network. A normal user might have accidentally performed an action which was not allowed. This uncertainty to the administrator entails to incomplete information in the game play. All the further interactions of administrator's ploy with the user is abstracted to take place in a single state $\xi_4$ of the main game in Figure 1. Here, the prime motivation of the administrator is to learn the "original" motivation of the user with the minimum number of interactions to identify the nature and motivation of the user and thus determine the suitable response. The administrator, based on his current belief about the user's nature and motivation, creates the situation in the state $\xi_4$. Administrator provides some actions as choices for the user during each interaction step, such that user's actions give the maximum information about his identity. The administrator has a finite set of such actions to offer the user to study his response. Offering them all at once is not advisable as it confuses the user making him change the original plan or become suspicious to altogether start behaving like a good user and may even try to log out inconspicuously. So, the actions should be provided as options at each step of this engagement to facilitate optimal user identification. Such a ploy can be characterized by the following structure to achieve this.

*Ploy*: *Formal description*: The administrator has a finite set of actions to offer to the user. Let this set be

$$A_{ploy} = \left\{ \alpha_{ploy_{j_0}}, \alpha_{ploy_{j_1}}, \alpha_{ploy_{j_2}}, \cdots, \alpha_{ploy_{j_x}} \right\} \tag{4}$$

where *x* is the total number of action-options that the administrator can offer at any general step *j* in the interaction. Based on the belief about the motivation of user at the time entering the $\xi_4$, the administrator associates a probability distribution over the set $A_{ploy}$, to represent the expected user's preference. $p_{\alpha_{ploy_j}}$ is the probability with which the user is expected to prefer to take the action $\alpha_{ploy_{i_j}}$ in the step *j* while facing the ploy. This implies

$$\sum_{i=1}^{x} p_{a_{ploy_{j_i}}} = 1 \tag{5}$$

Now the actions in the set $A_{ploy}$ are sorted in descending order by their associated probabilities.

$$A_{ploy_{sorted}} = \left\{ \alpha_{ploy_{i_0}}, \alpha_{ploy_{i_1}}, \cdots, \alpha_{ploy_{i_x}} \right\} \tag{6}$$

where

$$\left( p_{\alpha_{ploy_i}} \right) \leq \left( p_{\alpha_{ploy_j}} \right), \forall k \leq j. \tag{7}$$

Let the action set of the user at step 1 at $\xi_4$ be

$$A_{ploy_1}^1 = \left\{ \alpha_{ploy_{1_0}}, \alpha_{ploy_{1_1}}, \cdots, \alpha_{ploy_{1_y}} \right\}. \tag{8}$$

And in general at any given step *j* the user action set would be

$$A_{ploy_j}^1 = \left\{ \alpha_{ploy_{j_0}}, \alpha_{ploy_{j_1}}, \cdots, \alpha_{ploy_{j_y}} \right\}. \tag{9}$$

But this is essentially the similar looking action set as was in $\xi_1$ or $\xi_2$ or $\xi_3$ to start with from where the user landed in $\xi_4$. But the effect of the actions is different. As is true for any mixed strategy game, there exists a Nash strategy profile for the user and the administrator here too. This implies that there is an action guided by the user's Nash strategy, as identified/expected by the administrator, that the user is better off taking at this step. Let $\alpha_{ploy_{nash}}$ be that action in the step 1 in $\xi_4$. Administrator has to add/remove an action/actions at each step of interaction, to make the probability of the user taking actions as per Nash strategy, equiprobable with respect to user's current action set. This leads us to the following theorem.

*Theorem* 1: *In an incomplete information game, when the action prescribed by the Nash-strategy for a user is equiprobable with respect to other possible actions at a given state, the action selection is solely influenced by the original* *motivation of the player*[21]. *The player's selecting the prescribed action is a sufficient condition for the confirmation of the expected motivation (belief of payoff of user).*

*Proof:*

$$p_{\alpha^1_{ploy_{jnash}}} \cong \frac{\sum_{i=1}^{y} p_{\alpha_{ploy^1_{ji}}} - p_{\alpha^1_{ploy_{jnash}}}}{j_y - 1} \tag{10}$$

where $j_y$ is the number of actions in the current action set at step $j$. The main rationale is that the situation artificially created at the trapped state neither dissuade not encourage the user to take the action $\alpha^1_{ploy_{jnash}}$. Administrator cannot infer well if he makes the user do the action administrator choose for user. From the point of information to the user, he has as much information as he could with choice in background. The average information/Shannon-entropy given to him is at its maximum. As established in 3.5 we have

$$\left( I^1_{ploy_j} \right)_{max} = \sum_{i=1}^{y} p_{\alpha^1_{ploy_{ji}}} \log \left( 1 / p_{\alpha^1_{ploy_{ji}}} \right) \tag{11}$$

$$\square$$

where $I^1_{ploy_j}$ is the Average information over action set.

$$p_{\alpha^1_{ploy_{ji}}} = \frac{1}{j_i}, \forall j_i, 1 \leq j_i \leq y$$

Even though all the individual actions may not be equiprobable for user's selection, the main concern is regarding the action prescribed by the Nash strategy and its probability of being selected. From Equations (3.3) and (3.4), administrator has ensured that, if user takes the Nash action, then he is doing it at a maximum average information situation. This allows the sufficient confirmation to justify the previous belief. How ever if the user takes a different action then it also provides the direction for updating the belief. In either case, the administrator is not influencing the user's decisions and thus getting the most effective learning about user's original intention. The Equation (3.7) gives the ideal situation. But the minima of absolute value of the difference between LHS and RHS is to be considered for practical purpose. When the difference is positive then the administrator is dissuading the user and when it is negative he is encouraging the user from perusing the expected behavior. Both disturb the objective observation. Hence the absolute minimum difference as shown above is the best administrator can do.

## 10. Conclusion

An implementation of countering a real life attack is to be done to evaluate the dynamics and scopes for improvements. The Trap state provides a learning opportunity for the administrator while gaming with the user. The apposite game model would allow a Bayesian game with updating the belief about user, as the game progresses. The Bayesian Nash Equilibrium gives the useful solution concept to optimize the interaction. Exploring that with in this model would be the next step. The present model has static reward functions, which are generating the rewards in the three dimensional vector space. The next step would be to make it dynamic and learn during the game plays. This would make the model and resulting games truly generic to capture many more scenarios. A suitable method to compute the equilibria for the general cases and in particular security games is our planned future work.

---

[21] And not by any reactionary behavior towards other players.

## Acknowledgements

I want to acknowledge Anupama for the times during which this work was done.

# Appendix

## A. Example

This can be illustrated with an example. In the game in Section 5, an attacker trying to attack in $\xi_1$ may be affecting the eventual success of his attacking plans by revealing his attacking intention such that if he does not succeed with the present attempt, may later affect the reaction of the system administrator to promote to $\xi_2$ if he chooses $\alpha_{1_2}^1$. Thus information revealed to the other player about one's intention can affect the potential of the all next actions to propel the state transitions. Even if by chance, while the player is attacking, the game lands in the $\xi_2$ instead of $\xi_3$, he will now be under serious gaze of administrator. With this premise, logging out of the system may be better off with sign-out action, than to be stuck in Honeypot by the administrator. Thus the desirability of landing in $\xi_2$ after an attack action $\alpha_{1_3}^1$ may not be same as just landing with $\alpha_{1_2}^1$.

## B. Generalized Imperfect Information Factor

The model for 2-state game in [5] with the following extension can be used to calculate Nash equilibrium for an $N$-state game with $N < \infty$. From [22] we follow that the stochasticity of the game provides us with a probability $P^m(i, j)$ with which the state $\xi_j$ can occur from $\xi_i$ in $m^{th}$ iteration. The occurrence ratios $r_1, r_2, r_3$ & $r_4$ corresponding to the states $\xi_1, \xi_2, \xi_3$ & $\xi_4$ are given by

$$r_i = \lim_{m \to \infty} \frac{P(1, i) + P^2(1, i) + \cdots + P^m(1, i)}{m}, \ 1 \leq i \leq N \ \text{with} \ \sum_{i=1}^{N} r_i = 1, \text{and}$$

$N = 4$ for the game in Section 5. But the perceived occurrence ratios are different and is given by $r_i' = \sum_{j=1}^{N} p_{j_i}^k * r_j$. We note that the apparent strategy is different than the truly executing strategy. We have $\pi^{k'} = IIF * \pi^k$, where $IIF$ is called the *Imperfect Information Factor*. $IIF$ is defined as a matrix with entries $z_{ij}$ with $i$ rows and $j$ columns, with $z_{ij} = \dfrac{p_{j_i}^k * r_j}{r_i'}$. This gives the generalized $IIF$ for $N$ states. For the game in Section 5 with 5 states IIF is evaluated as shown below.

$$IIF = \left[ z_{ij} \right]_{5 \times 5} = \begin{bmatrix} \dfrac{p_{00}^k r_0}{r_0'} & \dfrac{p_{01}^k r_1}{r_0'} & \dfrac{p_{02}^k r_2}{r_0'} & \dfrac{p_{03}^k r_3}{r_0'} & \dfrac{p_{04}^k r_4}{r_0'} \\[2mm] \dfrac{p_{10}^k r_0}{r_1'} & \dfrac{p_{11}^k r_1}{r_1'} & \dfrac{p_{12}^k r_2}{r_1'} & \dfrac{p_{13}^k r_3}{r_1'} & \dfrac{p_{14}^k r_4}{r_1'} \\[2mm] \dfrac{p_{20}^k r_0}{r_2'} & \dfrac{p_{21}^k r_1}{r_2'} & \dfrac{p_{22}^k r_2}{r_2'} & \dfrac{p_{23}^k r_3}{r_2'} & \dfrac{p_{24}^k r_4}{r_2'} \\[2mm] \dfrac{p_{30}^k r_0}{r_3'} & \dfrac{p_{31}^k r_1}{r_3'} & \dfrac{p_{32}^k r_2}{r_3'} & \dfrac{p_{33}^k r_3}{r_3'} & \dfrac{p_{34}^k r_4}{r_3'} \\[2mm] \dfrac{p_{40}^k r_0}{r_4'} & \dfrac{p_{41}^k r_1}{r_4'} & \dfrac{p_{42}^k r_2}{r_4'} & \dfrac{p_{43}^k r_3}{r_4'} & \dfrac{p_{44}^k r_4}{r_4'} \end{bmatrix}$$

## C. $\epsilon$-Neighborhood Optimal Strategy Profile

The problem of finding the optimal exact Nash Equilibrium efficiently in the general case is challenging. Many progressive efforts are reported as discussed in Section 2. We plan to progress on those results in our future work. We here calculate an $\epsilon$-Approximate Equilibrium(AE) for an example of the stochastic game presented in 5. Traditionally, $\epsilon$-AE in a stochastic game is defined as the strategy profile, where no player can improve his expected average reward more than $\epsilon$ by unilateral strategy deviation. We introduce another variation $\epsilon$-Neighborhood Optimal Strategy Profile ($\epsilon$-**NOSP**) such that $0 \le |\epsilon| < 1$. The traditional definition of the $\epsilon$-ANE is dependent on the motivation being not greater than $\epsilon$ to deviate from the prescribed strategy profile for each of the players. When the game is stochastic, reward is valued at the average of rewards got by the execution of the game with the same strategy profile some number of times. We propose here a variation here due to two reasons.

1) Since the distribution of the rewards some times was bi-modal, the average may not be a comprehensive measure of the distribution.

2) There is a non-trivial probability of getting a higher reward for a player with a strategy with lower average reward than one with higher average reward, depending on their standard deviations.

Without assuming any distribution for the rewards over multiple game runs for a strategy profile, we calculate the $\epsilon$-NOSP $v^k_{\left(\pi^0, \pi^1, \pi^2, \cdots, \pi^k_i, \cdots, \pi^K\right)}$ with

$$\epsilon \ge \frac{v^k_{\left(\pi^0, \pi^1, \pi^2, \cdots, \pi^k_i, \cdots, \pi^K\right)_{\max}} - v^k_{\left(\pi^0, \pi^1, \pi^2, \cdots, \pi^k_i, \cdots, \pi^K\right)}}{\sigma\left(v^k_{\left(\pi^0, \pi^1, \pi^2, \cdots, \pi^k_i, \cdots, \pi^K\right)_{\max}}\right) + \sigma\left(v^k_{\left(\pi^0, \pi^1, \pi^2, \cdots, \pi^k_i, \cdots, \pi^K\right)}\right)}$$

where $\sigma$ is the standard deviation. This indicates the overlap of reward spectrum given by compared strategy profiles for the player *k*.